\newcommand{\docversion}{July 30, 2018}  
\documentclass{article}
\RequirePackage[l2tabu,orthodox,abort]{nag}
\usepackage[utf8]{inputenc}
\usepackage{authblk}
\usepackage[font=small]{caption}
\usepackage[parskip,repro]{reidpr}

\title{Deceptiveness of internet data\\ for disease surveillance}
\date{\docversion\ / LA-UR~17-24564}

\author[*$\S$]{Reid~Priedhorsky}
\author[*]{Dave~Osthus}
\author[*$\dagger$]{Ashlynn~R.~Daughton}
\author[*]{Kelly~R.~Moran}
\author[$\ddagger$]{Aron~Culotta}
\affil[*]{Los Alamos National Laboratory}
\affil[$\dagger$]{University of Colorado, Boulder}
\affil[$\ddagger$]{Illinois Institute of Technology}
\affil[$\S$]{corresponding author: \url{reidpr@lanl.gov}}

\graphicspath{{./}{fig/}{plot/}}

\DeclareSIUnit{\nothing}{\relax} 

\newtheoremstyle{assumption}%
{\abovedisplayskip}
{\belowdisplayskip}
{\setlength{\leftskip}{\parindent}\setlength{\rightskip}{\leftskip}}
{0pt} 
{} 
{.} 
{ } 
{\kern 1sp --- \itshape #1 #2} 

\theoremstyle{assumption}

\makeatletter
\newcommand{\Pa} [1][]{\def\tmp{#1}P\ifx\tmp\@empty\else^{#1}\fi}
\newcommand{\Pv} [1][]{P^{\text{v}#1}}
\newcommand{\Pnv}[1][]{P^{\text{nv}#1}}
\newcommand{\Pd} [1][]{P^{\text{d}#1}}
\newcommand{\Pnd}[1][]{P^{\text{nd}#1}}
\makeatother

\newcommand{\ot}{y}            
\newcommand{\qoi}{w}           
\newcommand{\QOI}{W}           
\newcommand{\sig}{u}           
\newcommand{\SIG}{U}           
\newcommand{\ns}{\delta}       
\newcommand{\NS}{\Delta}       
\newcommand{\nr}{\varepsilon}  
\newcommand{\NR}{\mathcal{E}}  
\newcommand{\f}{x}             
\newcommand{\F}{X}             
\newcommand{\Flen}{|\SIG|}     

\newcommand{\EN}  [1][]{\SIG#1}
\newcommand{\en}  [1][]{\sig#1}
\newcommand{\de}  [1][]{\hat\qoi#1}
\newcommand{\des} [1][]{\de^\text{s}#1}
\newcommand{\dens}[1][]{\de^\text{ns}#1}
\newcommand{\denr}[1][]{\de^\text{nr}#1}

\newcommand{\Err}{\mathbf{E}}
\newcommand{\riskg}{\mathbf{g}}
\newcommand{\risks}{\riskg^\text{s}}
\newcommand{\riskr}{\riskg^\text{r}}
\newcommand{\Riskg}{\mathbf{G}}

\begin{document}

\maketitle

\begin{abstract}


Quantifying how many people are or will be sick, and where, is a critical ingredient in reducing the burden of disease because it helps the public health system plan and implement effective outbreak response.
%
This process of \vocab{disease surveillance} is currently based on data gathering using clinical and laboratory methods; this distributed human contact and resulting bureaucratic data aggregation yield expensive procedures that lag real time by weeks or months. The promise of new surveillance approaches using internet data, such as web event logs or social media messages, is to achieve the same goal but faster and cheaper.
%
However, prior work in this area lacks a rigorous model of information flow, making it difficult to assess the reliability of both specific approaches and the body of work as a whole.

We model disease surveillance as a Shannon communication.
%
This new framework lets any two disease surveillance approaches be compared using a unified vocabulary and conceptual model. Using it, we describe and compare the deficiencies suffered by traditional and internet-based surveillance, introduce a new risk metric called \vocab{deceptiveness}, and offer mitigations for some of these deficiencies.
%
This framework also makes the rich tools of information theory applicable to disease surveillance.
%
This better understanding will improve the decision-making of public health practitioners by helping to leverage internet-based surveillance in a way complementary to the strengths of traditional surveillance.

\end{abstract}

\pagebreak
\section{Introduction}

%
%


Despite advances in medicine and public health, infectious disease still causes substantial morbidity and mortality~\cite{mathers2006}. \vocab{Disease surveillance} provides the data required to combat disease by identifying new outbreaks, monitoring ongoing outbreaks, and forecasting future outbreaks~\cite{horstmann1974}. However, traditional surveillance relies on in-person data gathering for clinical evaluations and laboratory tests, making it costly, difficult, and slow to cover the necessary large geographic areas and population. Disease surveillance using internet data promises to reach the same goals faster and at lower cost.

This promise depends on two things being true: (1)~people leave traces of their own and others' health status online and (2)~these traces can be extracted and used to accurately estimate disease incidence. Traces include search queries~\cite{ginsberg2008}, social media messages~\cite{culotta2013}, reference work usage~\cite{generous2014}, and blog posts~\cite{corley2010}.\footnote{One review of this body of research is our previous work~\cite{priedhorsky2017cscw}.}

The first claim is compelling, but the second is more elusive. For example, Google Flu Trends, a web system based on~\cite{ginsberg2008}, opened to great fanfare but proved to be inaccurate in many situations~\cite{lazer2014} and shut down in the summer of 2015~\cite{flutrendsteam2015}. The field has also encountered difficulty answering criticisms from the public health community on how it deals with demographic bias, media coverage of outbreaks, high noise levels, and other issues. Because the field's successes are based on observational studies in specific contexts, it is hard to know how robust or generalizable these approaches are or what unsuccessful alternatives remain unpublished.

\begin{figure}
  \includegraphics[width=\textwidth]{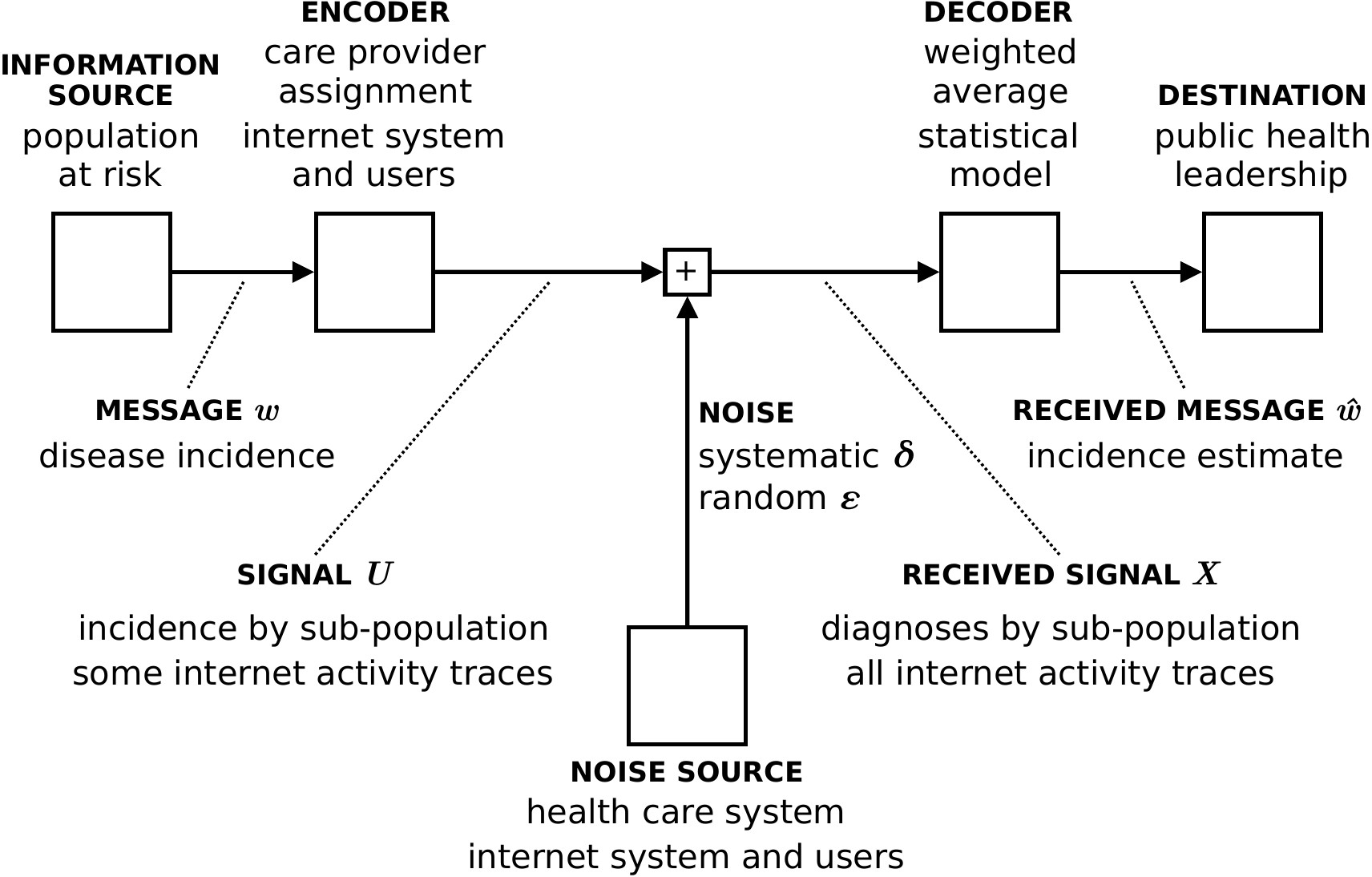}
  \caption{Schematic diagram of disease surveillance as a Shannon communication, patterned after Figure~1 of~\cite{shannon1948} and using the notation in this paper. This formulation lets one compare any two disease surveillance approaches with unified vocabulary and apply the rich tools of information theory.}
  \label{fig:schematic}
\end{figure}

\begin{figure}
  \includegraphics[width=\textwidth]{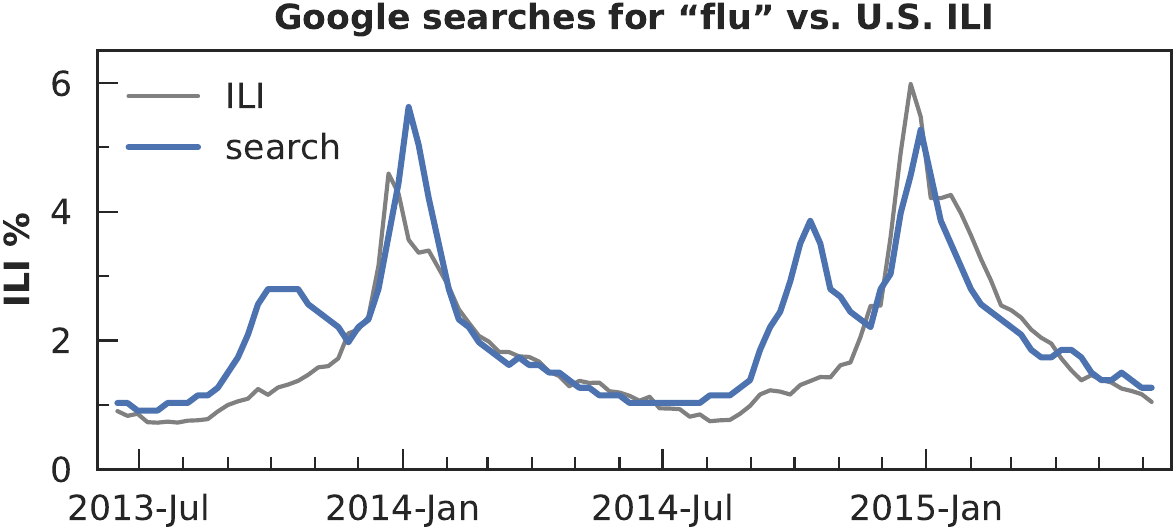}
  \caption{Google Trends U.S.\ search volume~\cite{google2017trends} for query string ``flu'' fitted to U.S.\ influenza-like-illness (ILI) data~\cite{cdc2017fluview}. The plot shows that search volume fails to track ILI early in the season, perhaps due to vaccination publicity schedules, demonstrating concept drift. We call this query \vocab{deceptive} during these periods.}
  \label{fig:concept-drift-google-flu}
\end{figure}
\begin{figure}
  \includegraphics[width=\textwidth]{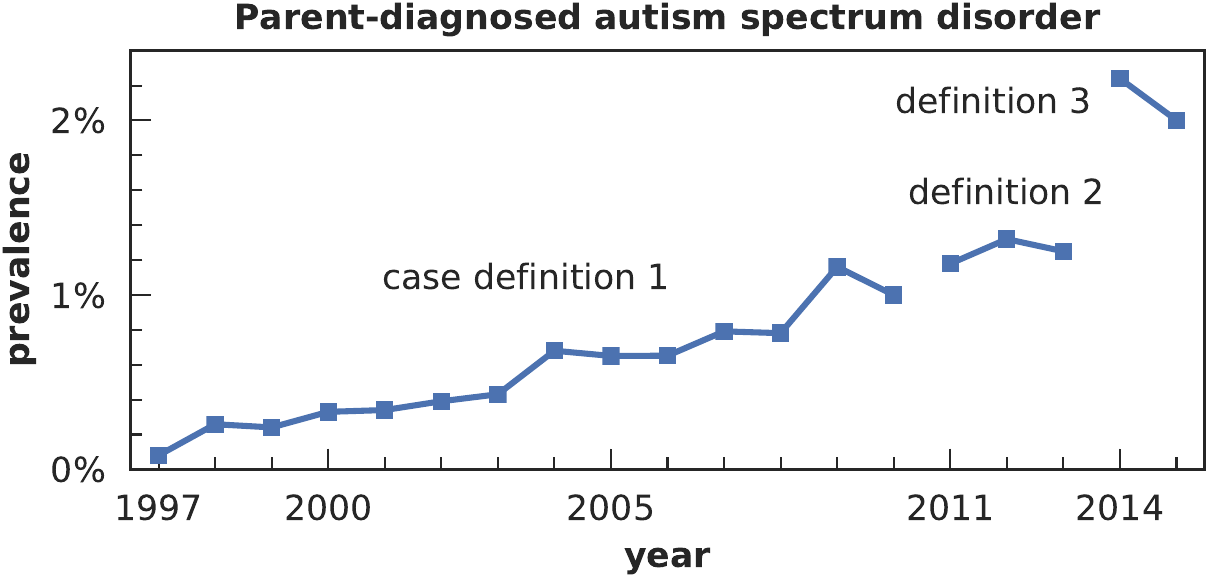}
  \caption{Prevalence of autism spectrum disorder as reported by parents has increased by a factor of~30 between 1997 and 2015 in a survey-based traditional surveillance approach~\cite{nhis2017,zablotsky2015}, but actual prevalence probably has not~\cite{zablotsky2015}. This plot shows drift both gradually as parental interpretations of diagnostic criteria change over time and suddenly when diagnostic criteria were revised.}
  \label{fig:concept-drift-autism}
\end{figure}

This paper argues that a necessary part of the solution is a mathematical model of the disease surveillance information flow. We offer such a model, describing both internet-based and traditional disease surveillance as a Shannon communication~\cite{shannon1948} (Figure~\ref{fig:schematic}). This lets us discuss the two approaches with unified vocabulary and mathematics. In particular, we show that they face similar challenges, but these challenges manifest and are addressed differently. For example, both face \vocab{concept drift}~\cite{gama2014}: Google search volume is sometimes, but not always, predictive of influenza incidence (Figure~\ref{fig:concept-drift-google-flu}), and traditional estimates of autism spectrum disorder are impacted by changing interpretations of changing case definitions~\cite{zablotsky2015} (Figure~\ref{fig:concept-drift-autism}).

Using this model, we introduce a new quality metric, \vocab{deceptiveness}, in order to quantify how much a surveillance system risks giving the right answer for the wrong reasons, i.e., to put a number on ``past performance is no guarantee of future results''. For example, basketball-related web searches correlate with some influenza seasons, and are thus predictive of flu, but should not be included in flu estimation models because this correlation is a coincidence~\cite{ginsberg2008}.

This approach lets us do three things. First, we show that neither approach is perfect nor has fully addressed its challenges. Second, we show that the two approaches are complementary; internet-based disease surveillance can add value to but not replace traditional surveillance. Finally, we identify specific improvements that internet-based surveillance can make, in order to become a trusted complement, and show how such improvements can be quantified.

In the body of this paper, we first describe a general mapping of disease surveillance to the components of Shannon communication, along with its challenges and evaluation metrics in terms of this mapping. We then cover the same three aspects of traditional and internet-based surveillance more specifically. We close with a future research agenda in light of this description.

\section{Disease surveillance in general}
\label{sec:all-surveillance}

Many aspects of traditional and internet-based disease surveillance are similar. This section lays out the bulk of our model in common and highlights the aspects that will be treated specifically later. We first describe disease surveillance as a Shannon communication, then use these concepts to discuss the challenges of surveillance and metrics to assess performance.

\begin{figure*}
  \includegraphics[width=\textwidth]{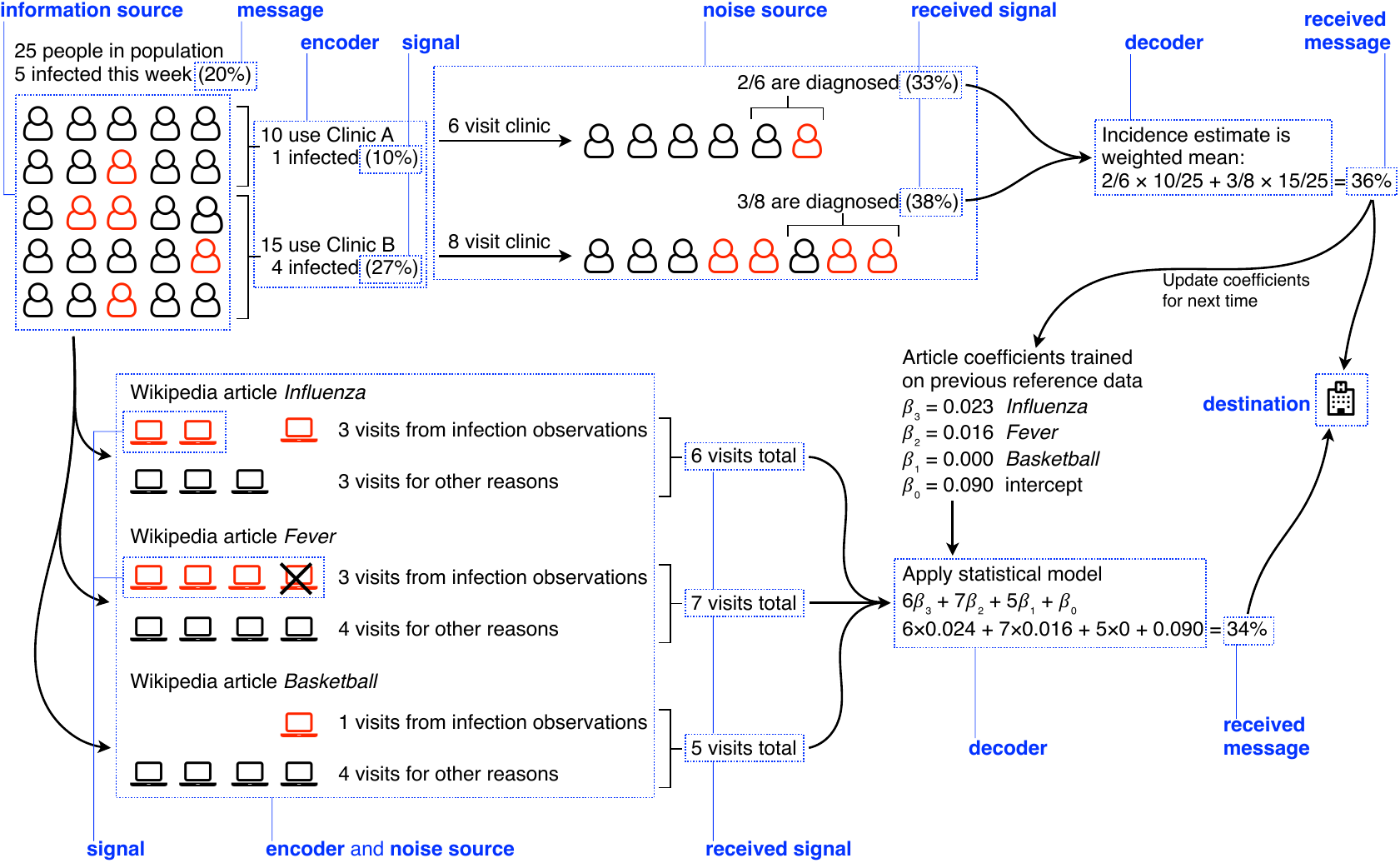}
  \caption{Example population of 25 people and how information about this population's disease status flows to public health decision makers. By traditional means (top flow), incidence information is encoded by the assignment of people to providers, added to noise introduced by the health care system, and decoded by a weighted average. By internet-based means (bottom flow), information is encoded and made noisy by Wikipedia use and decoded by a statistical model. (Note that encyclopedia article visits resulting from infection observations contain both signal and random noise, as illustrated by the extra visit to ``Influenza'' and the missed visit to ``Fever''.) Because quantitative internet-based surveillance needs reference data for training, it both depends on and reflects problems in traditional surveillance}
  \label{fig:flow-cartoon}
\end{figure*}

\subsection{As a Shannon communication}

Our model of disease surveillance as a Shannon communication is illustrated in schematic form in Figure~\ref{fig:schematic} and as an example population in Figure~\ref{fig:flow-cartoon}. This section walks through the key Shannon concepts and vocabulary, detailing how they fit a disease surveillance process.

The \vocab{information source} is a real-world phenomenon with a quantitative property we want to measure over some sequence of time intervals. Specifically, we assume that for each interval~$v$, there exists a well-defined \vocab{quantity of interest} (QOI)~$\qoi_v$.

In our disease surveillance context, we might ask, of United States residents (information source), what fraction become infected with influenza (QOI) during each week of the flu season (interval):\footnote{This is known as \vocab{incidence}. An alternate measure is \vocab{prevalence}, which is the number of active infections at any given time and which may be more observable by laypeople. Despite this, we illustrate our model with incidence because it is more commonly used by public health professionals. The two are roughly interchangeable unless the interval duration is considerably shorter than the typical duration of infection.}
\begin{equation}
  \qoi_v =     \frac{\text{people newly infected during } v}
                    {\text{total population at risk}}
\end{equation}

Each $\qoi_v$ comprises one \vocab{symbol}, and the \vocab{message} $\QOI$ is the sequence of all $\qoi_v$; i.e., $\QOI = \{ \qoi_1, \qoi_2, ..., \qoi_{|V|} \}$. Here, $\QOI$\! comprises the true, unobservable \vocab{epidemic curve}~\cite{nelson2013} that public health professionals strive for.

Next, each symbol is transformed by an \vocab{encoder} function to its corresponding \vocab{encoded symbol}, $\EN_v(\qoi_v)$. This transformation is distributed, so $\SIG_v$ is not a scalar quantity but rather a set of sub-symbols $\sig_{vi}$:
\begin{equation}
  \label{eq:encoderi}
  \EN_v(\qoi_v) = \{ \en_{v1}(\qoi_v),
                     \en_{v2}(\qoi_v),
                     \ldots ,
                     \en_{v\Flen}(\qoi_v) \}
\end{equation}
(Note that when clear from context, we use unadorned function names like $\SIG_v$ and $\sig_v$ to refer to their output, not the function itself, with the argument implied.)

For traditional disease surveillance, the encoder is the partitioning of the population at risk into sub-populations. For example, $\sig_{vi}$ might be the fraction of people served by clinic~$i$ who became infected during interval $v$.

For internet-based surveillance, the encoder is an internet system along with its users. Individuals with health concerns search, click links, write messages, and create other activity traces related to their health status or observations. For example, $\sig_{vi}$ might be the noise-free number of web searches for query $i$ motivated by observations of infection.

Both types of disease surveillance are noisy. Thus, $\SIG_v$ is unobservable
by the \vocab{decoder}. Instead, the \vocab{received symbol} at each interval is a set $\F_v$ of noisy features:
\begin{align}
  \label{eq:feature-noise}
  \f_{vi}(\qoi_v) &= \en_{vi}(\qoi_v) + \ns_{vi} + \nr_{vi} \\
  \label{eq:received-symbol}
  \F_v(\qoi_v) &= \{ \f_{v1}(\qoi_v),
                     \f_{v2}(\qoi_v),
                     \ldots ,
                     \f_{v\Flen}(\qoi_v) \}
\end{align}
Each observable feature $\f_{vi}$ is composed of unobservable signal $\sig_{vi}$, systematic noise $\ns_{vi}$, and random noise $\nr_{vi}$.

For traditional disease surveillance, the \vocab{noise source} is the front-line health care system along with individuals' choices on whether to engage with this system. For example, there is random variation in whether an individual chooses to seek care and systematically varying influence on those choices. Also, diagnostic criteria can be over- or under-inclusive, and providers make both random and systematic diagnosis errors. Considering these and other influences, $\f_{vi}$ might be the fraction of individuals visiting clinic $i$ who are diagnosed positive during interval $v$, which is not the same as the fraction of individuals in sub-population $i$ who become infected.

For internet-based surveillance, both the noise source and the encoder are in the internet system and its users. For example, whether a given individual with health observations makes a given search contains random variation and systematic biases; also, health-related articles are visited by individuals for non-health reasons. Technical effects such as caching also affect the data. $\f_{vi}$ might be the number of number of searches for query~$i$, which is different than the number of searches for query~$i$ motivated by observations of infection and unaffected by noise.

The decoder function $\de_v(\F_v)$ transforms the received symbol into an estimate of the QOI. For traditional disease surveillance, this is typically a weighted average of $\f_{vi}$; for internet-based surveillance, it is a statistical model trained against available traditional surveillance estimates, which lag real time.

Further details of encoding, noise, and decoding are specific to the different surveillance methods and will be explored below.

Finally, the \vocab{destination} of the communication is the people who act on the knowledge contained in $\hat\qoi_v$, in our case those who make public health decisions, such as public health professionals and elected or appointed leadership.

The problem is that $\hat\qoi_v \ne \qoi_v$ for both traditional and internet surveillance. This leads to the questions \emph{why} and \emph{by how much}, which we address next.

\subsection{Challenges}

QOI estimates $\hat\qoi_v$ are inaccurate for specific reasons.\footnote{This ontology of errors is related to that of survey statistics, which classifies on two dimensions~\cite{kalton1983}. \vocab{Sampling error} is \vocab{random} when the whole population is not observed and \vocab{systematic} when the selected sample is non-representative (selection bias). \vocab{Non-sampling error} is everything else, such as measurement and transcription error, and can also be random or systematic. } Feature $i$ might be non-useful because (a)~$\en_{vi}(\F_v)$ is too small, (b)~$\ns_{vi}$ is too large, (c)~the function $\en_{vi}$ is changing, and/or (d)~$\ns_{vi}$ is changing. These problems can be exacerbated by \vocab{model misspecification}, e.g. if the ``true'' relationship is quadratic but a linear model is used; we assume a reasonably appropriate model. The following sub-sections describe these four challenges.

\subsubsection{$\boldsymbol{\en_{vi}(\F_v)}$ too small: Signal to noise ratio}
\label{sec:missing-data}

Disease surveillance can miss useful information because the signal $\sig_{vi}$ is too small: either it does not exist ($\sig_{vi} = 0$) or is swamped by noise ($\sig_{vi} \ll \ns_{vi} + \nr_{vi}$).

For example, many diseases have high rates of asymptomatic infection; about half of influenza cases have no symptoms~\cite{cdc2015tables}. Asymptomatic infections are hard to measure because they do not motivate internet traces and infected individuals do not seek care. Thus, correction factors tend to be uncertain, particularly for specific outbreaks.

An example of the latter is diseases that have low incidence but are well-known and interesting to the public, such as rabies, which had two U.S. cases in 2013~\cite{adams2015}. These diseases tend to produce many internet activity traces, but few are related to actual infections.

\subsubsection{$\boldsymbol{\ns_{vi}}$ too large: Static sample bias}
\label{sec:sample-bias}

The populations sampled by disease surveillance differ from the population at large in systematic ways. For example, those who seek care at clinics, and are thus available to traditional surveillance, tend to be sicker than the general population~\cite{halzack2014}, and internet systems invariably have different demographics than the general population~\cite{duggan2015social}.


This problem increases systematic noise $\ns_{vi}$ for a given feature $i$ by the same amount in each interval $v$.


\subsubsection{$\boldsymbol{\sig_{vi}}$ changing: Encoding drift}
\label{sec:encoding-drift}

The encoder function can change over time, i.e., $\en_{vi} \ne \en_{ti}$ for different intervals $v \ne t$. For example, internet systems often grow rapidly; this increases the number of available observers, in turn increasing the average number of traces per infection.

This is a form of concept drift and reduces estimate accuracy when models erroneously assume function $\en_{vi} = \en_{ti}$.

\subsubsection{$\boldsymbol{\ns_{vi}}$ changing: Sample bias drift}
\label{sec:noise-drift}


Systematic noise can also change over time, i.e., $\ns_{vi} \ne \ns_{ti}$. A commonly cited cause is the ``Oprah Effect'' (named after the American television personality and businesswoman Oprah Winfrey), where public interest in an outbreak rises due to media coverage, in turn driving an increase in internet activity related to this coverage rather than disease incidence~\cite{butler2013}. Traditional surveillance can have similar problems, dubbed the ``Jolie Effect''; some scholars believe actress Angelina Jolie's well-publicized preemptive mastectomy caused a significant change in the population of women seeking breast cancer-related genetic tests~\cite{desai2016}.

This is another form of concept drift and reduces estimate accuracy when models erroneously assume $\ns_{vi} = \ns_{ti}$.

Some phenomena can be a mix of the two types of drift. For example, internet systems actively manipulate activity~\cite{lazer2014}, such as Google's search recommendations that update live while a query is being typed~\cite{google2016search}. This can lead users to create disease-related activity traces they might not have otherwise, whether motivated by an observed infection (encoding drift) or not (sample bias drift).


\subsection{Accuracy metrics}


Epidemiologists evaluate disease surveillance systems on a variety of qualitative and quantitative dimensions~\cite{thacker1988}. One of these is accuracy: how close is the estimated incidence to the true value? This can be considered from two perspectives. \vocab{Error} is the difference between the estimated and true QOI values, while \vocab{deceptiveness} is the fraction of an estimate that is based on coincidental rather than informative evidence. The latter is important because it quantifies the risk that future applications of a model will give less accurate estimates. This section defines the two metrics.

\subsubsection{Error}

We quantify error $\Err$ in the usual way, as the difference between the QOI and its estimate:
\begin{equation}
  \label{eq:error}
  \Err_v = \hat\qoi_v - \qoi_v
\end{equation}
Perfectly accurate estimates yield $\Err_v = 0$, overestimates $\Err_v > 0$, and underestimates $\Err_v < 0$. Importantly, $\Err_v$ is unobservable, because $\qoi_v$ is unobservable.

As we discuss below, traditional surveillance acknowledges that $\Err_v \ne 0$ but its methods often assume $\Err_v = 0$, and this finds its way into internet surveillance via the latter's training.

\subsubsection{Deceptiveness}



Our second metric, \vocab{deceptiveness}, addresses the issue that an estimate can be accurate for the wrong reasons. A deceptive search query is illustrated in Figure~\ref{fig:concept-drift-google-flu}. Another example for internet-based surveillance is the sport of basketball, whose season of play roughly coincides with the flu season. A naïve model can notice this correlation and build a decoder that uses basketball-related activity traces to accurately estimate flu incidence. Philosophers might say this model has a \vocab{justified true belief} --- it has a correct, evidence-based assessment of the flu --- but this belief is not knowledge because it is not based on \emph{relevant} evidence, creating a \vocab{Gettier problem}~\cite{gettier1963}. This flu model relies on a coincidental, rather than real, relationship~\cite{ginsberg2008}. We call such a model, based on deceptive features, also deceptive.

We quantify the deceptiveness of both disease incidence estimates and individual features for both specific intervals and the analysis period as a whole. For a specific interval, deceptiveness is the fraction of the feature or estimate's value that is the result of noise, and the deceptiveness for the analysis period as a whole is the maximum of any interval (or some other suitable aggregation function).

Specifically, the deceptiveness of a specific feature $i$ at interval $v$ is:\footnote{This formulation ignores the fact that a feature can have multiple, offsetting sources of noise. One could correct this problem by summing the absolute value of each noise source independently. In general, the definition of $\riskg$ should be adapted to specific applications.}
\begin{equation}
  \riskg_{vi} = \frac{|\ns_{vi}| + |\nr_{vi}|}
                     {|\sig_{vi}| + |\ns_{vi}| + |\nr_{vi}|}
                \in [0,1]
\end{equation}
One can also define systematic and random deceptiveness~$\risks$ and~$\riskr$ if the two types of noise should be considered separately.

We summarize the deceptiveness of a feature by its maximum deceptiveness over all intervals:
\begin{equation}
  \riskg_i = \max_v ( \riskg_{vi} )
\end{equation}

The deceptiveness of a complete estimate or model depends on the decoder function $\hat\qoi$, because a decoder that is better able to exclude noise (e.g., by giving deceptive features small coefficients) will reduce the deceptiveness of its estimates. Thus, we partition $\hat\qoi$ into functions that depend on signal $\des$, systematic noise $\dens$, and random noise $\denr$:
\begin{align}
  \NS_v       &= \{ \ns_{v1}, \ns_{v2}, ..., \ns_{v\Flen} \} \\
  \NR_v       &= \{ \nr_{v1}, \nr_{v2}, ..., \nr_{v\Flen} \} \\
  \de_v(\F_v) &= \des_v(\SIG_v) + \dens_v(\NS_v) + \denr_v(\NR_v)
\end{align}
The deceptiveness of an estimate $\Riskg$ is therefore:
\begin{align}
  \Riskg_v &=     \frac{|\dens_v(\NS_v)| + |\denr_v(\NR_v)|}
                       {|\des_v(\SIG_v)| + |\dens_v(\NS_v)| + |\denr_v(\NR_v)|}
              \in [0,1] \\
  \Riskg   &= \max_v ( \Riskg_v )
\end{align}
%

A deceptive feature is an example of a \vocab{spurious relationship}. These typically arise due to omitted variable bias: there is some confounding variable (e.g., day of year) that jointly influences both a deceptive feature and the QOI, creating a \vocab{model mis-specification}~\cite{angrist2009}. Corrections include adding confounds to the model as control variables and data de-trending. These effects can be captured by our approach in $\dens$ and $\denr$.

Like error, deceptiveness is unobservable, and more pernicious because quantifying it relies on the causality of the relationship between a feature and the QOI. This makes it easy and tempting to discount. Further, being a measure of risk, estimating it from past data requires care because relevant future outcomes cannot be incorporated. However, as we discuss below, additional information such as semantic relatedness can help with such estimates.

We turn next to a specific analysis of traditional disease surveillance as a Shannon communication.

\section{Traditional disease surveillance}

Traditional disease surveillance is based on in-person gathering of clinical or laboratory data. These data feed into a reporting chain that aggregates them and computes the corresponding QOI estimates.

This section completes the Shannon model for traditional surveillance by detailing the encoding, noise, and decoding steps. We then explore the implications of this model for the surveillance challenges and accuracy metrics.

\subsection{Shannon communication model}

As a concrete example to illustrate the encoding and decoding components of the traditional surveillance communication, we use the United States' seasonal influenza reporting system, ILInet, because it is one of the most featureful traditional surveillance systems. ILInet produces incidence estimates of \vocab{influenza-like-illness} (ILI): symptoms consistent with influenza that lack a non-flu explanation~\cite{cdc2016}. Other well-known surveillance systems include the European Surveillance System (TESSy)~\cite{nichols2014}, the Foodborne Diseases Active Surveillance Network (FoodNet)~\cite{henao2015}, and the United States' arboviral surveillance system (ArboNET)~\cite{lindsey2012}.

Data for ILInet is collected at roughly \num{2800} voluntarily reporting \vocab{sentinel provider} clinics~\cite{cdc2016}.  During each one-week interval, reporting providers count (a)~the total number of patients seen and (b)~the number of patients diagnosed with ILI. These numbers are aggregated via state health departments and the CDC to produce the published ILI estimates.


\begin{figure}
  \includegraphics[width=\textwidth]{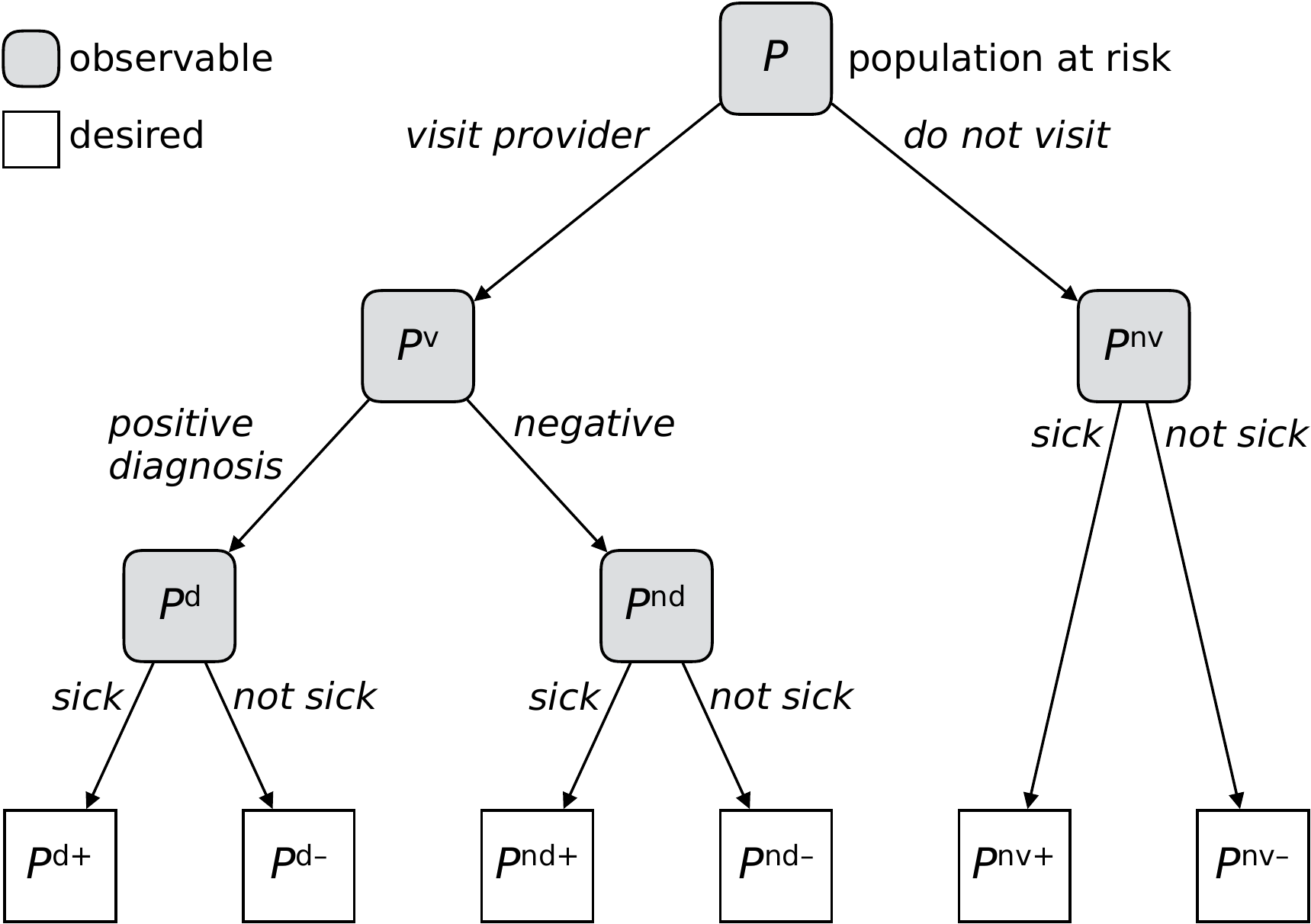}
  \caption{Hierarchical partition of the population at risk $\Pa$ into sets whose size is observable or not by traditional disease surveillance. This illustrates the mismatch between the information available to traditional surveillance and the quantity it is trying to measure.}
  \label{fig:ili-tree}
\end{figure}

This procedure suggests a hierarchical partition of the population at risk, as illustrated in Figure~\ref{fig:ili-tree}. The observable sets are:
\begin{itemize}

\item $\Pa_v$ : All members of the population at risk.

\item $\Pv_v$ : People who visit a sentinel provider.

\item $\Pnv_v$ : People who do not visit a sentinel provider.

\item $\Pd_v$ : People who visit a provider and meet the surveillance criteria, in this case because they are diagnosed with ILI.

\item $\Pnd_v$ : People who visit a provider but do not meet the surveillance criteria.

\end{itemize}
The sizes of these sets are observable: $|\Pa_v|$ from census data; $|\Pv_v|$, $|\Pd_v|$, and $|\Pnd_v|$ from provider reports; and $|\Pnv_v| = |\Pa_v| - |\Pv_v|$ by subtraction.

These sets are defined by provider visits and diagnosis, which is not the same as infection status. Thus, each is further partitioned into two unobservable subsets: those who acquire a flu infection during interval $v$ and those who do not (though they may have an active infection acquired during a previous interval). We denote these with superscript $+$ and $-$ respectively, e.g., $\Pa_v = \Pa[+]_v \cup \Pa[-]_v$; $\Pa[+]_v \cap \Pa[-]_v = \varnothing$.

We add a final, orthogonal partitioning into $\Flen$ sub-populations served by provider $i$. For example, the members of sub-population $i$ who visited a sentinel provider during interval $v$ and met the surveillance criteria comprise $\Pd_{vi}$. Each individual is a member of exactly one sub-population, and we assume for simplicity that members of sub-population $i$ visit either provider $i$ or none at all.\footnote{How individuals select providers does not affect our exposition. Real-world selections are typically driven by geography, socioeconomic status, demographics, and insurance.}

We can now model the encoding function that creates vector $\F_v$. Recall that the symbol that we wish to encode is the fraction of the population at risk that became infected during the interval:
\begin{align}
  \label{eq:trad-symbol}
  \qoi_v     &= \frac{|\Pa[+]_v|}{|\Pa_v|} \\
  |\Pa[+]_v| &= |\Pd[+]_v| + |\Pnd[+]_v| + |\Pnv[+]_v|
\end{align}
This can be partitioned by sub-population $i$:
\begin{align}
  \label{eq:trad-symbol-sub}
  \en_{vi}(\qoi_v) &= \frac{|\Pa[+]_{vi}|}{|\Pa_{vi}|} \\
  \label{eq:equal-pop}
         \beta_{i} &= \frac{|\Pa_{vi}|}{|\Pa_v|} \\
  \label{eq:trad-decode}
         \qoi_v    &= \sum_{i=1}^{\Flen} \beta_i \sig_{vi}
\end{align}

Note that $\qoi_v$ is defined in terms of infection status. On the other hand,  surveillance can observe only whether someone visits a sentinel provider and whether they are diagnosed with the surveillance criteria. Each feature $\f_{vi} \in \F_v$ is the fraction of patients visiting provider~$i$ who are diagnosed:\footnote{In reality, providers report the numerator and denominator of $\f_{vi}$ separately, but we omit this detail for simplicity.}
\begin{equation}
  \label{eq:trad-en-symbol}
  \f_{vi} = \frac{|\Pd_{vi}|}{|\Pv_{vi}|}
\end{equation}

Decoding to produce the estimate $\hat\qoi_v$ is accomplished as in Equation~\ref{eq:trad-decode}, using the population-derived $\beta$ as above:\footnote{This corresponds to the ILInet product \vocab{weighted ILI}, which weights by state population. A second product, plain \vocab{ILI}, gives each reporting provider equal weight. For simplicity, this paper uses unadorned \vocab{ILI} to refer to weighted ILI.}
\begin{equation}
  \de(X_v) = \sum_{i=1}^{\Flen} \beta_i \f_{vi}
\end{equation}

In short, the traditional surveillance communication is purpose-built to collect only disease-relevant information and thus couples with a simple decoder. Nonetheless, this communication does face the challenges enumerated above and is not noise-free. This we address next.

\subsection{Challenges}
\label{sec:trad-challenges}

This section details the four challenges outlined above in the context of traditional surveillance. The key issue is sample bias, both static and dynamic.

\subsubsection{Signal to noise ratio}

Traditional surveillance generally does not have meaningful problems with signal to noise ratio. While there are noise sources, as discussed below, these tend to be a nuisance rather than a cause of misleading conclusions.

This is due largely to the purpose-built nature of the surveillance system, which works hard to measure only what is of interest. To avoid missing cases of important rare diseases, where signal to noise ratio is of greatest concern, monitoring is more focused. For example, any provider who sees a case of rabies must report it~\cite{adams2015}, in contrast to ILInet, which captures only a sample of seasonal flu cases.

\subsubsection{Static sample bias}

Patient populations ($\Pv$) differ systematically from the general population ($\Pa$). For example, patients in the Veteran Health Administration tend to be sicker and older than the general population~\cite{halzack2014}, and traditional influenza systems tend to over-represent severe cases that are more common in children and the elderly~\cite{savard2015}.

Sampling bias can in principle be mitigated by adjusting measurement weights to account for the ways in which the populations differ, for example by age~\cite{savard2015}, triage score~\cite{savard2015}, and geography~\cite{cdc2016}. This, however, requires quantitative knowledge of the difference, which is not always available or accurate.

ILInet uses the number of sites that report each week, state-specific baselines, and age~\cite{cdc2016}. For influenza, age is particularly important for mitigating sample bias, because severity and attack rate differ widely between age groups~\cite{savard2015}.

\subsubsection{Encoding drift}

In our model, traditional surveillance is immune to encoding drift because the encoder functions contain only $\qoi_v$ (Equation~\ref{eq:trad-symbol-sub}); there are no other parameters to effect drift.

\subsubsection{Sample bias drift}

Traditional surveillance is designed to be consistent over time, but it does not reach this goal perfectly and is not immune to drifting sample bias.

One source of drift is that case definitions change. Figure~\ref{fig:concept-drift-autism} illustrates this for autism spectrum disorder, and the definition of ILI has also varied over time~\cite{jiang2015}. Another is that individuals' decisions on whether to seek care drift; for example, media coverage of an outbreak~\cite{rubin2010} or a celebrity editorial~\cite{desai2016} can affect these decisions. These and others do not affect the newly infected and non-newly-infected equally, so the sample bias and associated systematic noise changes.


This suggests that traditional surveillance might be improved by better addressing sample bias. Bias could be reduced by sampling those not seeking care, e.g., by random telephone surveys of the general population for flu symptoms, and it could be better corrected with improved static and dynamic bias models.

\subsection{Accuracy}

Traditional surveillance features include patients not newly infected ($\Pd[-]_{vi} \subset \Pd_{vi}$) and exclude people who are newly infected but receive a negative diagnosis or do not visit a sentinel provider ($\Pnd[+]_{vi} \cap \Pd_{vi} = \varnothing$ and $\Pnv[+]_{vi} \cap \Pv_{vi} = \varnothing$, respectively). Causes include lack of symptoms, diagnosis errors by individual or provider, and access problems. That is:
\begin{equation}
                 \f_{vi}
  =              \frac{|\Pd_{vi}|}{|\Pv_{vi}|}
  \ne \frac{|\Pa[+]_{vi}|}{|\Pa_{vi}|}
  =              \sig_{vi}
\end{equation}

In practice, this problem is addressed by essentially assuming it away. That is, consumers of traditional surveillance assume that the fraction of patients  diagnosed with the surveillance criteria does in fact equal the new infection rate among the general population:
\begin{align}
  \label{eq:infected-visitors}
  \frac{|\Pd_{vi}|}{|\Pv_{vi}|}  &= \frac{|\Pa[+]_{vi}|}{|\Pa_{vi}|} \\
                        \f_{vi}  &= \sig_{vi}
\end{align}
Some recent work extends this to assume that the quantities are proportional, independent of $v$~\cite[e.g.]{shaman2013}.

Epidemiologists know this assumption is dubious. They are careful to refer to ``ILI'' rather than ``influenza'', do not claim that ILI is equal to incidence, and give more consideration to changes over time rather than specific estimate values. Under these caveats, the measure serves well the direct goals of surveillance, such as understanding when an outbreak will peak and its severity relative to past outbreaks~\cite{cdc2016}.

Both error and deceptiveness in traditional surveillance depend on noise. This can be quantified by rewriting Equation~\ref{eq:trad-en-symbol}:
\begin{align}
  \f_{vi}             &=   \sig_{vi}
                         + \frac{|\Pd_{vi}|}{|\Pv_{vi}|}
                         - \frac{|\Pa[+]_{vi}|}{|\Pa_{vi}|} \\
  \ns_{vi} + \nr_{vi} &=   \frac{|\Pd_{vi}|}{|\Pv_{vi}|}
                         - \frac{|\Pa[+]_{vi}|}{|\Pa_{vi}|}
\end{align}
As one might expect, the noise in feature $\f_{vi}$ is the difference between the fraction of clinic-visiting, diagnosed patients and the fraction of the general population newly infected.

One challenge is separating the noise into systematic and random components. For example, if a newly infected person chooses to seek care, how much of that decision is random chance and how much is systematic bias? The present work does not address this, but one can imagine additional parameters to quantify this distinction.

\subsubsection{Error}

Recall that error is the difference between the true QOI and its estimate:
\begin{equation}
  \Err_v = \hat\qoi_v - \frac{|\Pa[+]_v|}{|\Pa_v|}
\end{equation}

The unobservable ratio $\frac{|\Pa[+]_v|}{|\Pa_v|}$ can be estimated, but with difficulty, because one must test the entire population, not just those who seek care or are symptomatic. One way to do this is a \vocab{sero-prevalence study}, which looks for antibodies in the blood of all members of a population; this provides direct evidence of past infection rates~\cite{gibbons2014}.

The adjustments discussed above in \S\ref{sec:trad-challenges} also reduce error by improving $\hat\qoi_v$.

%

\subsubsection{Deceptiveness}

One can write the deceptiveness of a specific feature as follows:
\begin{equation}
  \riskg_{vi} = \frac{\left| \frac{|\Pd_{vi}|}{|\Pv_{vi}|} - \sig_{vi} \right|}
                     {\sig_{vi} + \left|   \frac{|\Pd_{vi}|}{|\Pv_{vi}|}
                                         - \sig_{vi} \right|}
\end{equation}
Note that because we only know $\ns_{vi}$ and $\nr_{vi}$ as a sum, we must use $|\ns_{vi} + \nr_{vi}|$ rather than $|\ns_{vi}| + |\nr_{vi}|$, making the metric less robust against offsetting systematic and random error.

This again depends on an unobservable ratio, meaning that in practice it must be estimated rather than measured.

$\qoi_v \neq \hat\qoi_v$ in traditional disease surveillance due to unavoidable systematic and random noise. Nevertheless, it frequently provides high-quality data of great social value. For example, the distinction between a case of influenza and a case of non-influenza respiratory disease with flu-like symptoms is often immaterial to clinical resource allocation and staffing decisions, because the key quantity is volume of patients rather than which specific organism is at fault. $\hat\qoi_v$ provides this. Thus, traditional surveillance is an example of a situation where the quantity of interest $\qoi_v$ is elusive, but a noisy and biased surrogate $\hat\qoi_v$ is available and sufficient for many purposes. Accordingly, traditional surveillance has much in common with internet-based surveillance, as we next explore.

\section{Internet-based disease surveillance}

In contrast to traditional surveillance, which is based on data gathered by direct observation of individual patients, internet-based disease surveillance conjectures that traces of people's internet activity can drive a useful incidence estimation model. Thus, its Shannon communication differs, as do the corresponding implications for its challenges and accuracy.

\subsection{Shannon communication model}

Internet-based surveillance leverages found features optimized for unrelated purposes,\footnote{For example, looking at illustrations of cats~\cite{rathergood2010}.} instead of purpose-built ones as in traditional surveillance.

Rather than diagnosis counts made by health care providers, features in internet surveillance arise by counting defined classes of activity traces; thus, they are much more numerous. For example, Wikipedia contains 40~million encyclopedia articles across all its languages~\cite{zachte2016statistics}, versus \num{2800} ILInet providers. As a corollary, while traditional features offer useful information by design, internet-based features are the result of complex guided and emergent human behavior and thus overwhelmingly non-informative. That is, most internet features provide no useful information about flu incidence, but a few might. For example, the number of times the Wikipedia article ``Influenza'' was requested in a given week plausibly informs a flu estimate, while ``Basketball'' and ``George H.~W.\ Bush'' do not.

Raw traces such as access log lines are converted into counts, such as the number of requests for each article, searches for each query string, or n-gram mentions per unit time. In the case of Wikipedia, $\f_{vi} \in \mathbb{Z} \ge 0$ is the number of requests for Wikipedia article $i$ during interval $v$.

These trace counts comprise the features. For example, for each new infection during interval $v$, some number of requests $\sig_{vi}$ for every Wikipedia article $i$ is generated. Added to this signal is systematic noise $\ns_{vi}$ (article requests made for other reasons), and random noise $\nr_{vi}$ (random variation in article traffic). Thus, the features observable to the decoder are the familiar mix of signal and noise:
\begin{equation}
  \tag{\ref{eq:feature-noise}}
  \f_{vi} = \sig_{vi} + \ns_{vi} + \nr_{vi}.
\end{equation}

Internet features use a different unit than traditional features. For traditional surveillance, the symbol and feature units are both the person, who has a specific infection status and can be assigned a positive or negative diagnosis. For internet surveillance, the unit is \vocab{activity traces}. These include traces of \vocab{information seeking} activity such as Wikipedia article requests~\cite{generous2014} or Google searches~\cite{ginsberg2008} as well as \vocab{information sharing} activity such as n-grams in Twitter messages~\cite{culotta2013}. This mismatch between feature units (trace counts) and encoded symbol units (people) is a challenge for internet-based surveillance that is not shared with traditional surveillance.

Like traditional surveillance, the decoder is a function of the observable data (features). Often, this is a parameterized model, such as a linear model, that contains parameters that must be estimated. This is done by fitting the model against a reference data set, which is usually incidence estimates from a traditional surveillance system. We denote these reference data as $\ot_v$. For a linear model, the fit is set up as:
\begin{equation}
  \label{eq:linear-fit}
  \ot_v = \sum_{i=1}^{\Flen} \beta_i \f_{vi} + \beta_0 + \nr_v
\end{equation}

The slopes $\beta_i$ are estimated in two steps. The first is a filter: because including the very large number of irrelevant internet features in a regression would be noisy and inefficient, most of them are removed by setting $\beta_i = 0$. This procedure might be a context analysis (i.e., all features semantically unrelated to the outbreak of interest are removed), a correlation analysis (features where the correlation between $\ot$ and $\f_i$ is below a threshold are removed), or something else. The goal is to greatly reduce the number of features, e.g. from 40~million to tens or hundreds in the case of Wikipedia, implementing the modelers' prior belief over which features are relevant and not deceptive.

Next, $\beta_0$ and the remaining $\beta_i$ are estimated by fitting the collection of $\f_i$ to $\ot$ using some kind of regression. These estimated coefficients constitute a fitted model that can be applied to a previously unseen set of features $\F_t$, producing an estimate of the reference $\hat\ot_v$ and in turn the QOI $\hat\qoi_v$.

This illustrates how internet-based disease surveillance is inextricably bound to traditional surveillance. Quantitative statistical estimates require quantitative reference data, and traditional surveillance in some form is what is available. Internet surveillance thus inherits many of traditional surveillance's flaws; for example, if the reference data are biased, the internet estimates will share this bias.

However, internet estimates do have important advantages of timeliness and cost because the information pipeline, once configured, is entirely automated. While for contexts like U.S.\ ILI, which has a lag of 1--2 weeks~\cite{cdc2016}, this speed advantage is modest, in other contexts traditional surveillance lags by months or an entire season, a critical flaw~\cite{jajosky2004}.

\subsection{Challenges}

Internet-based disease surveillance faces a new set of challenges derived from its use of a non-purpose-built encoder.

\subsubsection{Signal to noise ratio} This is a significant problem for internet-based surveillance. For example, the English Wikipedia article ``Rabies'' was requested 1.6~million times in 2013~\cite{henrik2014}, but there were only two U.S.\ cases that year~\cite{adams2015}. Thus, it's certain that $\sig \ll \ns + \nr$ and plausible that $\sig = 0$, making requests for this article a deceptive feature.

This limits the applicability of internet-based surveillance. Outbreaks likely occur having \emph{no} features with sufficiently low deceptiveness, making them impossible to surveil reliably with internet data. This may be true even if fitting yields a low-error model, because fitting seeks out correlation rather than causality; recall the basketball problem.

Models can address this by estimating deceptiveness and down-weighting features with high deceptiveness. Thus, accurately estimating deceptiveness is important. We explore this further below.

\subsubsection{Static sample bias}


Internet surveillance suffers from sample bias in multiple ways. First, disease-related activity traces in a given internet system reflect people's responses to the real world, including reflections of their own health status or that of people they observe (who are likely to be similar to them). Internet use varies systematically based on factors related to health status, such as age and other demographics; in turn, activity traces vary systematically by those same factors. For example, social media tends to be modestly skewed on most demographic variables studied, including gender, age, race, education, income, and urban/rural~\cite{duggan2015social}.

Second, the unit mismatch between the message (people) and signal (trace counts) can increase this bias. For example, one might hypothesize that a younger person with respiratory symptoms might post more messages about the situation than an older person with the same symptoms. If this is systematically true, then the the number of activity traces is additionally biased on top of the user bias noted above. This opportunity for bias is not shared with traditional surveillance.

However, the fitting process compensates for these biases. For example, if men are overrepresented among a systems' users and this is evident in their activity traces, the gendered traces will be down-weighted, correcting for the overrepresentation. Thus, the resulting estimates will be roughly as biased as the reference data.


\subsubsection{Encoding drift}

This issue can be a problem for internet surveillance, because the assumption that $\en_{vi} = \en_{ti}$ for all intervals $v$ and $t$ (Equation~\ref{eq:linear-fit}) is false. For example, modern search sites use positive feedback: as more people search for flu, the system notices and offers more suggestions to search for flu~\cite{lazer2014}, thus increasing the number of searches per infection.

Models can address this by incorporating more realistic assumptions about how $\en$ changes over time, for example that it varies smoothly.

\subsubsection{Sample bias drift}

This is perhaps the most pernicious problem for internet surveillance. That is, the assumption that $\ns_{vi} = \ns_{ti}\ \forall\, v,t$ (also Equation~\ref{eq:linear-fit}) is simply false, sometimes very much so. Causes can include gradual user and activity pattern drift over time as well as sudden ``Oprah Effect'' shifts.

Models can address this by treating $\ns$ over time more realistically. Opportunities include explicit model terms such as smooth variation, leveraging deceptiveness estimates to exclude deceptive features, and including data sources to estimate current sample bias, though taking advantage of such opportunities well may be challenging.

We argue that in order for any internet-based surveillance proposal to be credible, it must address signal-to-noise ratio and drift issues using careful, quantitative, and application-aware methods. On the other hand, static sample bias can be addressed by simply following best practices. We propose some assessment options in the next section.

\subsection{Accuracy}

Error for internet-based models can be easily computed against the reference data $\ot$, but because the QOI $\qoi$ is not observable even after the passage of time, true error is not observable either. This has two consequences. First, any imperfections in the traditional surveillance are not evident in the error metric; thus, it is prudent to evaluate against multiple reference data. Second, simulation studies where $\qoi$ is known are important for properly evaluating a proposed approach.

Deceptiveness adds an additional problem: it depends not only on the quantitative relationship between features and the QOI but their causal relationship as well. This makes it fundamentally unknowable, even with perfect reference data.

However, the situation is not quite so grim, as proxies are often available. For example, our previous work showed that limiting Wikipedia models to semantically related articles reduced error~\cite{priedhorsky2017cscw}. Models can also reduce risk by monitoring carefully for the breakdown in coincidental correlation that makes deceptive features troublesome. Our point is not that the situation is hopeless but that that effective internet-based models must grapple quantitatively with deceptiveness.

\begin{figure}
  \includegraphics[width=\textwidth]{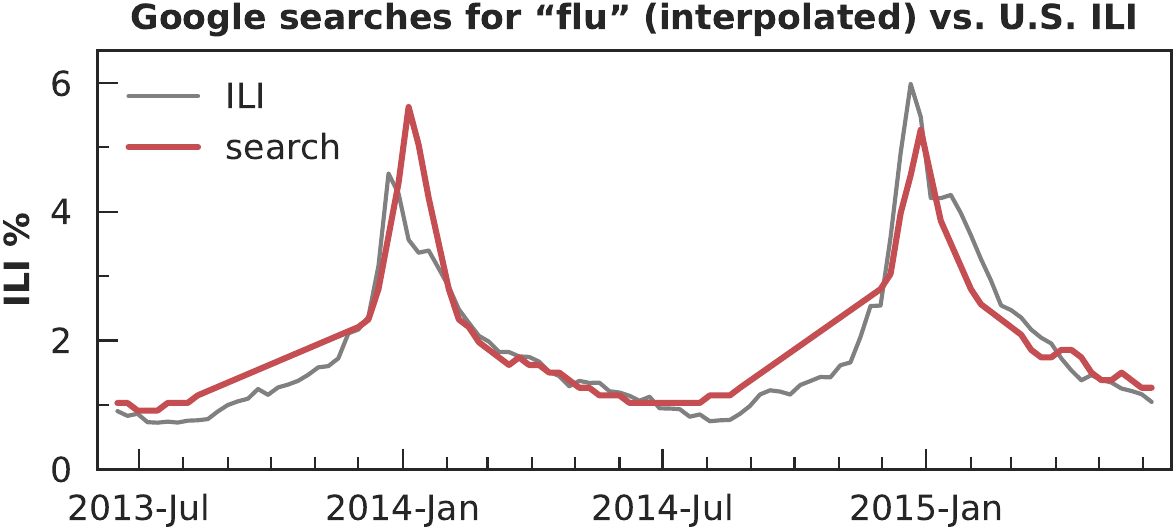}
  \vspace{0pt} \\
  \includegraphics[width=\textwidth]{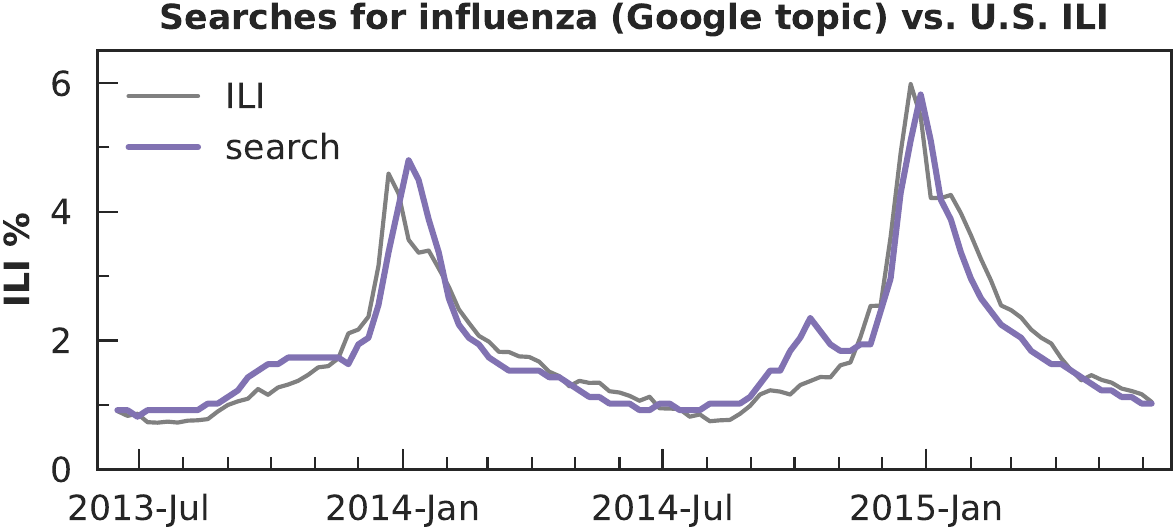}
  \caption{Alternatives to the deceptive feature in Figure~\ref{fig:concept-drift-google-flu}~\cite{cdc2017fluview,google2017trends}. These examples show that additional information can sometimes be used to reduce deceptiveness in internet features. At top, expert knowledge about the patterns of seasonal flu is used to identify and remove the deceptive period, replacing it with a linear interpolation. At bottom, Google's proprietary machine learning algorithms are used to identify and count searches likely related to the disease, rather than occurrences of a raw search string.}
  \label{fig:less-drift-google-flu}
\end{figure}

We argue that doing so successfully is plausible; compare the features in Figure~\ref{fig:less-drift-google-flu} to the one in Figure~\ref{fig:concept-drift-google-flu}. Properly designed internet-based surveillance might be almost as accurate as traditional surveillance but available faster, providing similar value in a more nimble, and thus actionable, fashion.

\section{Discussion}

Traditional disease surveillance is not perfect. It has problems with static sample bias and, to a lesser degree, drifting bias. These are consequences of the fact that traditional surveillance can only gather information from people in contact with the health care system. Thus, expanding access to health care, outreach, and further work on compensating for these biases can improve incidence estimates produced by traditional surveillance.

Internet-based surveillance is not perfect either. In addition to inheriting most of the problems of traditional surveillance, with which it is necessarily entangled for training data, it adds signal-to-noise problems, further sample bias issues, and further drift as a consequence of its found encoder. Yet because it is inexpensive and more timely, it holds great promise. So what is to be done? We propose three things.

\begin{enumerate}

\item Because deceptiveness matters, model features should be selected that have both high correlation with reference data (to drive good predictions) \emph{and} low deceptiveness (to reduce the risk of including coincidentally predictive features). Existing data fitting techniques address the first part, and we recommend adding a deceptiveness estimation step using additional information such as semantic relatedness.

\item Because past model performance by itself is not predictive of future performance, model proposals should also include an analytical argument supporting their performance. One approach is to (a)~define application-appropriate bounds on acceptable estimate error and deceptiveness, (b)~derive the properties of input features necessary to meet those bounds, and (c)~show that input features with those properties are available.

\item Because performance metrics include quantities unobservable in the real world, no model should be evaluated only on real data. Models should also be evaluated on simulated disease outbreaks, which do not replicate the real world in every detail but are valid tools for understanding~\cite{lee2008}, where all quantities are known.

\end{enumerate}

Internet-based disease surveillance cannot replace traditional surveillance, but it can be an important complement because it is faster and cheaper with broader reach. By better understanding the deficiencies of both along with their commonalities and relationship, using the tools proposed above, we can build quantitative arguments describing the added value of internet-based surveillance in specific situations. Only then can it be relied on for life-and-death public health decisions~\cite{generous2014}.

\section*{Acknowlegements}

This work was supported in part by the U.S.\ Department of Energy through the LANL/LDRD Program. A.C.\ was funded in part by the National Science Foundation under awards IIS-1526674 and IIS-1618244. Sara Y.\ Del Valle, John Szymanski, and Kyle Hickmann provided important feedback.

\bibliographystyle{plainurl}
\bibliography{refs}

\end{document}